\documentclass[aps,pre,twocolumn,showpacs]{revtex4}
\usepackage{epsfig}
\usepackage{times}
\begin{document}

\title{Group-size effects on the evolution of cooperation in the spatial public goods game}

\author{Attila Szolnoki$^1$ and Matja{\v z} Perc$^2$}
\affiliation
{$^1$Research Institute for Technical Physics and Materials Science,
P.O. Box 49, H-1525 Budapest, Hungary \\
$^2$Faculty of Natural Sciences and Mathematics, University of Maribor, Koro{\v s}ka cesta 160, SI-2000 Maribor, Slovenia}

\begin{abstract}
We study the evolution of cooperation in public goods games on the square lattice, focusing on the effects that are brought about by different sizes of groups where individuals collect their payoffs and search for potential strategy donors. We find that increasing the group size does not necessarily lead to mean-field behavior, as is traditionally observed for games governed by pairwise interactions, but rather that public cooperation may be additionally promoted by means of enhanced spatial reciprocity that sets in for very large groups. Our results highlight that the promotion of cooperation due to spatial interactions is not rooted solely in having restricted connections amongst players, but also in individuals having the opportunity to collect payoffs separately from their direct opponents. Moreover, in large groups the presence of a small number of defectors is bearable, which makes the mixed phase region expand with increasing group size. Having a chance of exploiting distant players, however, offers defectors a new way to break the phalanx of cooperators, and even to resurrect from small numbers to eventually completely invade the population.
\end{abstract}

\maketitle

The public goods game \cite{nowak_06, sigmund_10} is played in groups and captures the essential social dilemma in that collective and individual interests are inherently different. Players must decide simultaneously whether they wish to contribute to the common pool, \textit{i.e.} to cooperate, or not. All the contributions are then multiplied to take into account synergetic effects of cooperation, and the resulting amount is divided equally among all group members irrespective of their strategies. Selfish players obviously should decline to contribute if the investment costs exceed the return of the game. However, if nobody decides to invest the group fails to harvest the benefits of a collective investment, and the society may evolve towards the ``tragedy of the commons'' \cite{hardin_g_s68}. Yet despite of the obvious social dilemma, observations indicate that individuals cooperate much more in public goods games than expected \cite{fehr_ars07}, which calls for the identification of mechanisms that can sustain cooperation. The sustenance of cooperation in sizable groups of unrelated individuals, as is the case by the public goods game, is particularly challenging since group interactions tend to blur the trails of those who defect. Unlike by pairwise interactions, reciprocity \cite{axelrod_84, nowak_n98} often fails as it is not straightforward to determine with whom to reciprocate. Social enforcement, on the other hand, may work well, although it is challenged by the fact that it is costly (see \cite{sigmund_tee07} for a review).  Recently studied ways of promoting cooperation in public goods games include the introduction of volunteering \cite{hauert_s02, semmann_n03} and the introduction of social diversity by means of complex interaction networks \cite{santos_n08, zhang_hf_epl11}, random exploration of strategies \cite{traulsen_pnas09}, as well as various forms of reward \cite{rand_s09, hauert_jtb10, szolnoki_epl10, hilbe_prsb10} and punishment \cite{hauert_s07, helbing_ploscb10, sigmund_n10, szolnoki_pre11}, to name but a few.

Spatial reciprocity \cite{nowak_n92b}, being part of the big five \cite{nowak_11},
is long established as a prominent mechanism for the evolution of cooperation \cite{nowak_s06}. The spatial public goods game \cite{brandt_prsb03} in particular, is interesting also from the viewpoint of physics, for example in terms of phase transitions \cite{szabo_prl02}, pattern formation \cite{wakano_pnas09}, effects of inhomogeneous player activities \cite{guan_pre07}, diversity \cite{yang_hx_pre09} and noise \cite{szolnoki_pre09c}, as well as coevolutionary processes \cite{wu_t_epl09} and processes taking place on complex networks \cite{gomez-gardenes_chaos11,galan_pone11}. While the efficiency of spatial reciprocity is known to be vitally affected by the structure of interaction graphs \cite{szabo_pr07}, there is still a lack of studies systematically analyzing the impact of group size on the evolution of cooperation. Although it is traditionally assumed that very large groups should result in mean-field behavior due to the emergence of all-to-all coupling (see \textit{e.g.} \cite{szabo_pre09}), certain studies suggest that this may not always be the case \cite{wu_t_pre09}. Adding to this the experimental findings \cite{isaac_jpe94}, indicating that larger groups (of size $40$ or $100$) provide public goods more efficiently than small groups (of size $4$ or $10$), clearly outlines the need for clarifying the importance of the group size, especially for games that are governed by group interactions.

\begin{figure}
\centerline{\epsfig{file=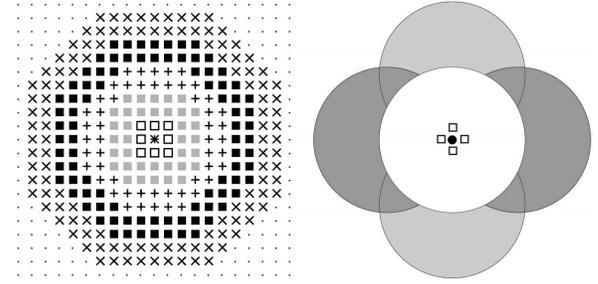,width=7.7cm}}
\caption{Left: Schematic presentation of different group sizes $G$ on the square lattice. Depicted are groups containing $G=9$ (open square), $45$ (grey square), $97$ (plus), $185$ (filled square) and $301$ (cross) players, respectively. The focal player is marked by a star. Right: Schematic presentation of possible sources of strategy invasion. Three different cases are considered, namely, the adoption can only be attempted from the nearest neighbors (open squares), from the focal group (white patch), or from all the groups (grey areas) where the focal player is a member.}
\label{scheme}
\end{figure}

Here we therefore study the evolution of cooperation in the public goods game on the square lattice, whereon initially each player on site $x$ is designated either as a cooperator ($s_x = C$) or defector ($s_x = D$) with equal probability.
We note, however, that the main findings do not depend on the host lattice topology because the large group size interactions diminish the fine topological differences. Players can collect payoffs from groups ranging in size from $G=5$ to $445$, as depicted schematically in Fig.~\ref{scheme}. In addition, we also consider different groups of players that are eligible to act as strategy donors, ranging from nearest neighbors only to all players that are members in the groups containing the focal player, \textit{i.e.} the one potentially adopting a new strategy. Note that there exist exactly $n=G$ groups containing any given player $x$ (one group where player $x$ is focal and $n-1$ groups where this is not the case). Each selected player $x$ acquires its payoff $P_x$ by accumulating its share of the public good from all the $n$ groups with which it is affiliated (unless stated otherwise). Without loss of generality cooperators contribute $1$ to the pool while defectors contribute nothing, and subsequently all the contributions within a group are multiplied by the enhancement factor $r$ and divided equally amongst all the members. Employing the Monte Carlo simulation procedure, each elementary step involves randomly selecting one focal player $x$ and one player $y$ that is eligible to act as a strategy donor. Following the accumulation of payoffs $P_x$ and $P_y$ as described above, player $y$ tries to enforce its strategy $s_y$ on player $x$ in accordance with the probability $W(s_y \rightarrow s_x)=\{1+\exp[(P_x-P_y)/K]\}^{-1}$, where $K$ determines the uncertainty by strategy adoptions \cite{szolnoki_pre09c}. To account for the different number of groups affecting the absolute values of the payoffs when increasing $G$ (and thus indirectly influencing $W$), parameters $r$ and $K$ must be considered properly normalized with $G$ to ensure relevant comparisons of results.
During a Monte Carlo step (MCS) all players will have a chance to pass their strategy once on average. For the results presented below we used the square lattice having $L=400$ to $1600$ linear size and up to $10^7$ MCS before determining the stationary fraction of cooperators $\rho_C$ within the whole population.

\begin{figure}
\centerline{\epsfig{file=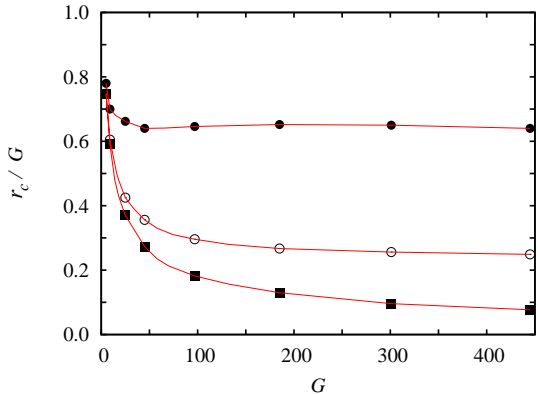,width=8cm}}
\caption{Critical multiplication factor $r_c$ in dependence on $G$. Strategy donors were selected only from the four nearest neighbors (filled squares), from within the group of players where player $x$ is focal (open circles), or amongst all the groups where player $x$ is a member (filled circles). The uncertainty by strategy adoptions was $K/G=0.1$ (the normalization of $K$ with $G$ takes into account the number of groups participating in the accumulation of payoffs).}
\label{whole}
\end{figure}

Figure~\ref{whole} features the critical multiplication factor $r_c$ at which cooperators die out in dependence on $G$. Above this $r_c$ value the cooperators can coexist with defectors by forming a mixed phase. If exceeding a second critical $r_c$ value (not shown here), the defectors will die out and the system will arrive at the pure $C$ phase, as demonstrated in previous works considering small group sizes \cite{szolnoki_pre09c, shi_dm_pa09, wu_t_pre09, lei_c_pa10, liu_rr_pa10, peng_d_epjb10, zhang_cy_epjb11}. In the present work we focus on the group-size dependence of the lower critical $r_c$ that limits the surviving chance of cooperator strategy. As Figure~\ref{whole} suggests, increasing the group size can drastically decrease the minimally required $r$ for the sustenance of cooperation, and there is no indication of arriving at mean-field behavior (note that $r_c=G$ in the well-mixed case \cite{hauert_tpb08}) even for very large groups. However, the positive effect depends significantly on the available set of potential strategy donors. The smaller the latter (nearest neighbors $\to$ focal group $\to$ all groups), the stronger the promotion of cooperation induced by large $G$. An alternative, and in fact more interesting, interpretation is that the larger the difference between interaction (used for the accumulation of payoffs) and replacement (used for selecting potential strategy donors) groups, the smaller the $r_c$ at any given $G$. This is different from what was reported in \cite{ohtsuki_prl07} for games governed by pairwise interactions, where cooperators were found diminishing as the overlap between interaction and replacement graphs was lessened.

\begin{figure}
\centerline{\epsfig{file=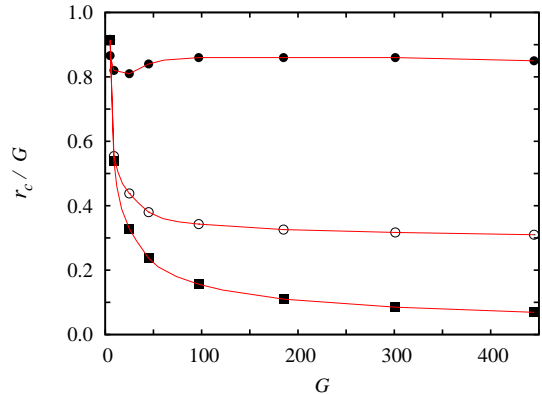,width=8cm}}
\caption{Critical multiplication factor $r_c$ in dependence on $G$, as obtained when the payoffs are acquired from a single group where the corresponding players are focal. The symbols correspond to those used in Fig.~\ref{whole}, determining the set of potential strategy donors. The uncertainty by strategy adoptions was $K=0.1$ (note that the normalization with $G$ is unnecessary since all the payoffs originate from a single group).}
\label{focal}
\end{figure}

Our observations can be corroborated further by considering public goods games where each player $x$ acquires its payoff $P_x$ only from the one group where it is focal. Figure~\ref{focal} shows the results. The most relevant difference with the results presented in Fig.~\ref{whole} can be observed for the case where strategy donors are selected amongst all the groups where player $x$ is a member (filled circles).
Note that distant players can interact indirectly here, \textit{i.e.} although they do not collect payoffs from the same group, their strategies influence the income of the other player. In this case there exists an optimal group size where $r_c$ is minimal (instead of a continuous downward trend), although the well-mixed limit ($r_c=G$), and in fact even the small-group limit ($r/G \to 0.915$), is never reached for very large $G$. Results presented in Fig.~\ref{focal} lead to the conclusion that it is beneficial for the evolution of cooperation not only if the interaction and replacement groups are different, but also, when players have the ability to play the game (collect their payoffs) with other players who are beyond the scope of potential donors of a new strategy. For the uppermost curve in Fig.~\ref{focal} (closed circles) this is not warranted (note that the payoffs are collected only from the group where a given player $x$ is central, while strategy donors are sought from all the groups where player $x$ is member), and it is indeed there where the promotion of cooperation by means of large groups is least effective. Nevertheless, large groups are definitely better suited for the effective provision of public goods under unfavorable conditions (small $r$) then small groups, thus supporting the experimental findings of Isaac et al. \cite{isaac_jpe94}.

\begin{figure}
\centerline{\epsfig{file=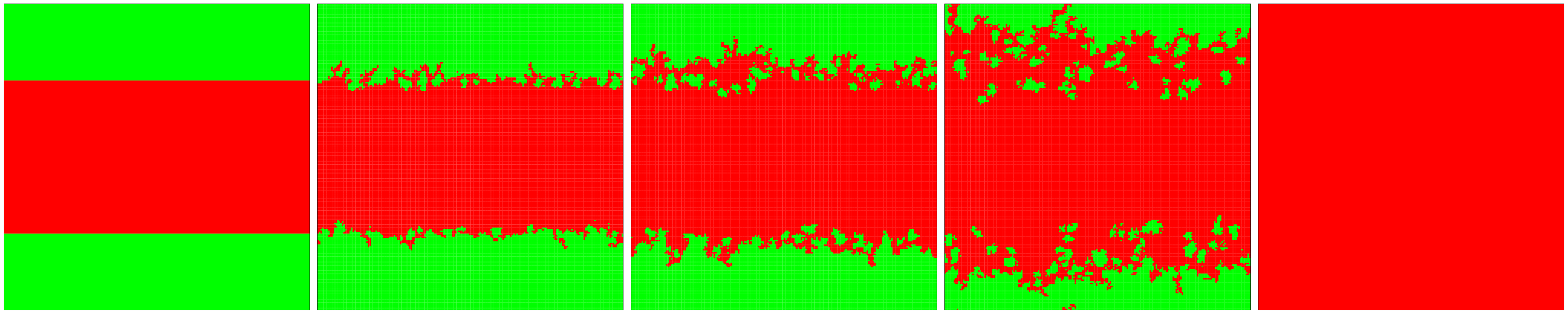,width=8.5cm}}
\centerline{\epsfig{file=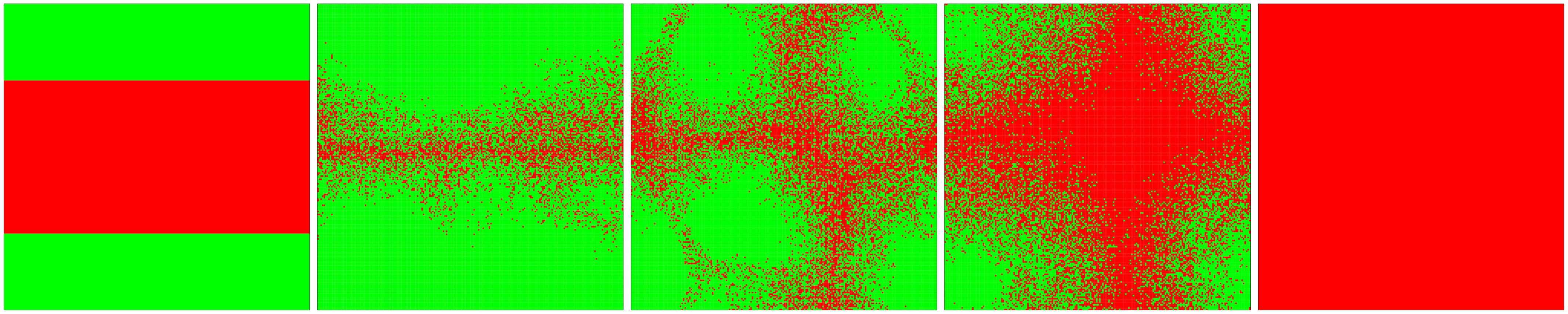,width=8.5cm}}
\caption{(Color online) Characteristic snapshots of the spatial grid for $G=5$ (top row) and $G=301$ (bottom row), using focal group imitation range as obtained for $K/G=0.1$ and $L=200$ system size. Cooperators are marked by green (light grey) and defectors by red (dark grey) colors. In both cases the final outcome is a full $D$ phase where normalized synergy factors are almost equal far from the transition points ($r/G=0.74$ and $r/G=0.24$, respectively). Prepared initial states were used to highlight the two significantly different strategy invasion processes. Snapshots in the top row were taken at MCS $=0, 50, 100, 200$ and $1500$, while in the bottom row they were taken at MCS $=0, 30, 100, 300$ and $400$.}
\label{snaps}
\end{figure}

Characteristic snapshots of the spatial grid for small and large $G$, as depicted in Fig.~\ref{snaps} in the top and bottom row, respectively, serve well to understand the differences in the evolutionary process that is brought about by differently sized groups. For small groups ($G=5$, top row), the evolution of strategies proceeds with the characteristic propagation of the fronts of the more successful strategy (in this case $D$) until eventually the maladaptive strategy $C$ goes extinct. For intermediate values of $r$, we would observe the well-known clustering of cooperators \cite{nowak_n92b}. On the other hand, for large groups ($G=301$, bottom row) the cooperator clusters are very strong and can easily outperform the defectors, even if $r$ is very small. However, as the number of defectors in the large groups goes down, their payoff suddenly becomes very competitive, to the point where defectors can strike back and invade the seemingly invincible cooperative clusters. Such an alternating time evolution is completely atypical and was previously associated with cooperators only (see for example \cite{perc_bs10}), \textit{i.e.} the density of cooperators typically goes down initially, until some form of reciprocity or a feedback effect establishes itself and enables the cooperators to win back lost ground to defectors. For spatial public goods games played in large groups we here demonstrate that the scenario is exactly the opposite. Defectors are the ones who can resurrect from small numbers to overtake cooperators, and it is indeed the difficulty of prevention of this negative backfiring of the initial cooperative success that limits the success of large groups to sustain cooperation at even smaller multiplication factors.

\begin{figure}
\centerline{\epsfig{file=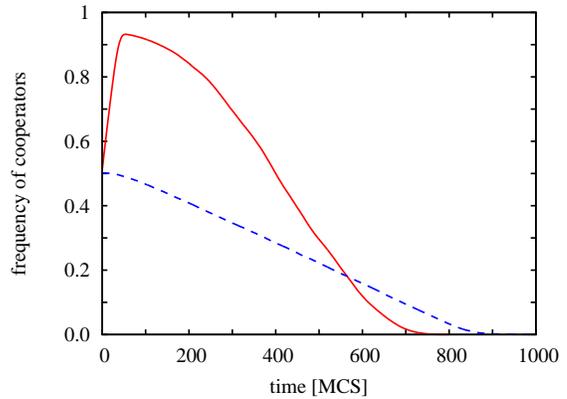,width=8cm}}
\caption{(Color online) Time courses of the density of cooperators $\rho_C$ for $G=5$ (dashed blue line) and $G=445$ (solid red line), starting from the mentioned prepared initial state (see Fig.~\ref{snaps}) and using all-group imitation range, as obtained for $K/G=0.1$ and $L=800$ system size. Synergy factors were $r/G=0.7$ and $r/G=0.6$, respectively.}
\label{time}
\end{figure}

The two opposite time courses presented in Fig.~\ref{time} illustrate the atypical evolutionary process at large $G$ succinctly. To obtain smooth curves, we have used larger system sizes ($L=800$) and averaged the data over $50$ independent runs. While for small $G$ (dashed line) the fraction of cooperators $\rho_C$ decreases monotonically to zero, the outlay for $G=445$ (solid line) is very much different. There we can first observe a significant {\it increase} in $\rho_C$, which is brought about by the formidably strong cooperative phalanx, which can easily defeat weak defectors deep in the $D$ domain. The dissolution of $D$ domains, however, serves well the surviving defectors who then become the ``leaders'' of a counter attack that eventually leads to the complete extinction of cooperators. Hence, we can observe the fall of $\rho_C$, although as emphasized, this one is due to completely different circumstances than the one reported for the $G=5$ case. The time evolution of defectors, as we have demonstrated for $G=445$ (dominance following near extinction), was previously associated with cooperative behavior only, and it is only the special impact of distant invasion, which is made possible by large groups, on the evolution of cooperation that is able to offer such a reversal of expected roles of the two strategies.

In sum, we have studied the evolution of cooperation in the spatial public goods games on the square lattice, thereby focusing on revealing the impact of different group sizes on the effective provision of public goods. Motivated by the experimental findings indicating that larger groups are advantageous to small groups \cite{isaac_jpe94}, we find that large groups indeed significantly promote the evolution of cooperation. Quite remarkably, if only the interaction and replacement groups are sufficiently different, and if players have the ability to play the public goods game with at least some of the players that are then not considered as potential donors of a new strategy, the large groups prove impervious to defectors even at very low multiplication factors. Since spatial reciprocity is inherently routed in the formation of compact cooperative clusters, it seems natural that larger groups, potentially giving rise to larger cooperative clusters, will be more effective in warranting high levels of cooperation than small groups. However, it is the size of large groups that may backfire on the cooperators when the number of defectors in such groups become very low. Then the advantages of defection become so strong that cooperators may still be defeated despite of their stellar start. It is mainly this mechanism that limits the success of large groups to sustain cooperation and puts a lid on the pure number-in-the-group effect \cite{isaac_qje88}. We would also like to emphasize that the identified mechanism of promotion of cooperation by means of participation in large groups is robust and independent of details such as the uncertainty by strategy adoptions or the local structure of the interaction network. In particular, the joint membership in large groups will indirectly link vast numbers of players \cite{szolnoki_pre09c}, thus rendering local as well global structural properties of interaction networks practically irrelevant for the final outcome of the game. There are several examples, like local and federal tax payment, health insurance or pension systems, when people are involved in partly separated large structured common ventures. Without applying our model directly to such systems, the present work offers an explanation why in fact cooperation can survive even when the benefits of large-scale collaboration are relatively modest.

\begin{acknowledgments}
Authors acknowledge support from the Hungarian National Research Fund (grant K-73449), the Bolyai Research Fund, and the Slovenian Research Agency (grant Z1-2032).
\end{acknowledgments}

\end{document}